\documentclass[10pt]{revtex4}
\usepackage{amssymb,epsf,epsfig,graphics}
\usepackage{latexsym}
\begin{document}

\title{Rotating black strings in $f(R)$-Maxwell theory}
\author{A. Sheykhi$^{1,2}$\footnote{asheykhi@shirazu.ac.ir}, S. Salarpour $^{3}$ and Y. Bahrampour $^{4}$\footnote{bahram@uk.ac.ir}}
\address{$^1$ Physics Department and Biruni Observatory, College of
Sciences, Shiraz University, Shiraz 71454, Iran\\
 $^2$ Research Institute for Astrophysics and Astronomy of Maragha
(RIAAM), P.O. Box 55134-441, Maragha, Iran \\
        $^3$ Department of Physics, Shahid Bahonar University, P.O. Box 76175, Kerman, Iran\\
       $^4$ Department of Mathematics, Shahid Bahonar University, P.O. Box 76175, Kerman, Iran  }
\begin{abstract}
In general, the field equations of $f(R)$ theory coupled to a
matter field are very complicated and hence it is not easy to find
exact analytical solutions. However, if one considers traceless
energy-momentum tensor for the matter source as well as constant
scalar curvature, one can derive some exact analytical solutions
from $f(R)$ theory coupled to a matter field. In this paper, by
assuming constant curvature scalar, we construct a class of
charged rotating black string solutions in $f(R)$-Maxwell theory.
We study the physical properties and obtain the conserved
quantities of the solutions. The conserved and thermodynamic
quantities computed here depend on function $f'(R_{0})$ and differ
completely from those of Einstein theory in AdS spaces. Besides,
unlike Einstein gravity, the entropy does not obey the area law.
We also investigate the validity of the first law of
thermodynamics as well as the stability analysis in the canonical
ensemble, and show that the black string solutions are always
thermodynamically stable in $f(R)$-Maxwell theory with constant
curvature scalar. Finally, we extend the study to the case where
the Ricci scalar is not a constant and in particular $R=R(r)$. In
this case, by using the Lagrangian multipliers method, we derive
an analytical black string solution from $f(R)$ gravity and
reconstructed the function $R(r)$. We find that this class of
solutions has an additional logaritmic term in the metric function
which incorporates the effect of the $f(R)$ theory in the
solutions.

\end{abstract}
\maketitle

\newpage

\section{Introduction\label{Intro}}
There has been considerable attentions in the past years in modified
gravity theories, specially $f(R)$ theory which is one of the
encouraging candidates for explaining the current accelerating of
the universe expansion \cite{Odin,Capo} (see also \cite{Anto} for a
comprehensive review on $f(R)$ theories). In fact $f(R)$ theories
can be regarded as the simplest extension of general relativity.
Many $f(R)$ models have passed all the available experimental tests
and fit the cosmological data. To prevent a ghost state, $f'(R)>0$
for $R \geq R_0$ is required \cite{Nun,Fara}. $f''(R)>0$ for $R \geq
R_0$, is needed to avoid the negative mass squared of a scalar-field
degree of freedom (tachyon) \cite{Anto}. $f(R)\rightarrow
R-2\Lambda$ for $R \geq R_0$, is required for the presence of the
matter era and for consistency with local gravity constraints
\cite{Anto}. It was shown that $f(R)$ theories can be considered as
general relativity with an additional scalar field that provide new
insight in the two cases of Brans-Dicke theory with ${\omega}_0 = 0$
and ${\omega}_0 = -3/2$ \cite{Soti}.

There have been a lot of works in the literature attempting to
construct static and stationary black hole solutions in $f(R)$
gravity theories. One may expect that some signatures of black
holes in $f(R)$ theories will be in disagreement with the expected
physical results of Einstein's gravity. In \cite{Cruz} the authors
studied general solutions in $f(R)$ theory using a perturbation
approach around the Einstein-Hilbert action. In \cite{Psal} black
hole solutions were found by adding dynamical vector and tensor
degrees of freedom to the Einstein-Hilbert action. Also, the
transition from neutron stars to a strong scalar- field state in
$f(R)$ gravity has been studied in \cite{Nova}. Physical
properties of the matter forming an accretion disk in the
spherically symmetric background in $f(R)$ theories were explored
in \cite{Pun}. In Ref. \cite{Lobo} the construction of traversable
wormhole geometries was discussed in $f(R)$ gravity. The
Schwarzschild-de Sitter black hole like solutions of $f(R)$
gravity were obtained for a positively constant and a non-constant
curvature scalar in \cite{Cogn} and \cite{Sebas}, respectively. A
black hole solution was obtained from $f(R)$ theories by requiring
the negative constant curvature scalar \cite{Cruz}. If $1 +
f'(R_0)
> 0$, this black hole is similar to the Schwarzschild-AdS (SAdS)
black hole. It was argued that $f(R)$ and SAdS black holes have no
big difference in thermodynamic quantities when using the
Euclidean action approach and replacing the Newtonian constant $G$
by $G_{\rm eff} = G/(1 + f'(R_0))$ \cite{Cruz}. It is also
interesting to study black hole solutions in $f(R)$ theory coupled
to a matter field. In general, the field equations of $f(R)$
theory coupled to the matter field are very complicated and hence
it is not easy to find exact analytical solutions. In order to
construct the constant curvature scalar black hole solutions from
$f(R)$ gravity coupled to the matter, the trace of its
energy-momentum tensor $T_{\mu \nu}$ should be zero \cite{Moon}.
Two examples for the traceless $T_{\mu \nu}$ are Maxwell and
Yang-Mills fields which were studied in \cite{Moon,Habib}.
Thermodynamics and properties of these solutions were also studied
in ample details \cite{Moon}. It was found that these solutions
are similar to the Reissner-Nordstr\"{o}m--AdS (RNAdS) black hole
when making appropriate replacements \cite{Moon}. The Kerr-Newman
black hole solutions with non-zero constant scalar curvature in
$f(R)$-Maxwell theory, their thermodynamics, as well as their
local and global stability were also studied in \cite{Alex}.

In this paper we would like to continue the investigation on the
$f(R)$ black holes, by constructing a new class of charged
rotating black string solutions in $R+f(R)$-Maxwell theory with
constant  curvature scalar. The traceless property of the
energy-momentum tensor of the Maxwell field plays a crucial role
in our derivation. With assumptions $R_0<0$ and
$1+f^{\prime}(R_0)>0$ our solution is similar to charged black
string solution in AdS space with suitable replacing the
parameters. We will also suggest the suitable counterterm which
removes the divergences of the action. We calculate the conserved
and thermodynamic quantities of these black strings by using the
counterterm method. We obtain a Smarr-type formula for the mass of
the black string and check the validity of the first law of
thermodynamics. We perform the stability analysis in the canonical
ensemble and show that the black strings are always
thermodynamically stable in $f(R)$-Maxwell theory with constant
curvature scalar. Finally, we extend the study to the case where
the Ricci scalar is not constant and in particular $R=R(r)$ and
derive an analytical black string solution.

\section{Field Equations and solutions\label{Field}}
We start from the four-dimensional $R+f(R)$ theory coupled to the
Maxwell field
\begin{eqnarray}
I_{G} &=&-\frac{1}{16\pi }\int_{\mathcal{M}}d^{4}x\sqrt{-g}\left(
R+f(R)-F_{\mu \nu }F^{\mu \nu }\right)-\frac{1}{8\pi
}\int_{\partial \mathcal{M}}d^{3}x\sqrt{-h }\Theta (h ),
\label{Act}
\end{eqnarray}
where ${R}$ is the Ricci scalar curvature, $F_{\mu \nu }=\partial
_{\mu }A_{\nu}-\partial _{\nu }A_{\mu }$ is the electromagnetic
field tensor, and $A_{\mu }$ is the electromagnetic potential. The
last term in Eq. (\ref{Act}) is the Gibbons-Hawking boundary term.
It is required for the variational principle to be well-defined.
The factor $\Theta$ represents the trace of the extrinsic
curvature for the boundary ${\partial \mathcal{M}}$ and $h$ is the
induced metric on the boundary. The equations of motion can be
obtained by varying the action (\ref{Act}) with respect to the
gravitational field $g_{\mu \nu }$ and the gauge field $A_{\mu }$
which yields the following field equations
\begin{eqnarray}
{R}_{\mu \nu } \left(1+f^{\prime}(R)\right)-\frac{1}{2}g_{\mu
\nu}(R+f(R))+\left(g_{\mu\nu}\nabla ^{2}-\nabla _{\mu} \nabla
_{\nu}\right)f^{\prime}(R)=8 \pi T_{\mu \nu} ,  \label{FE1}
\end{eqnarray}
\begin{eqnarray}
\nabla_{\mu } F^{\mu \nu}=0,  \label{FE2}
\end{eqnarray}
with the energy-momentum tensor
\begin{equation}
T_{\mu \nu}=\frac{1}{4 \pi}\left(F_{\mu \eta }F_{\nu }^{\text{
}\eta }-\frac{1}{4}g_{\mu \nu }F_{\lambda \eta }F^{\lambda \eta
}\right). \label{T}
\end{equation}
The above energy-momentum tensor is traceless in four dimension,
i. e., $T^{\mu}_{\text{ } \ \mu}=0$. As we mentioned already this
property plays an important role in our derivation. In Eq.
(\ref{FE1})  the ``prime'' denotes differentiation with respect to
curvature scalar $R$. Assuming the constant curvature scalar
$R=R_0$, the trace of Eq. (\ref{FE1}) yields
\begin{eqnarray}
R_0\left(1+f'(R_0)\right)-2\left(R_0+f(R_0)\right)=0,
\end{eqnarray}
Solving the above equation for negative $R_0$, gives
\begin{eqnarray}
R_0=\frac{2f(R_0)}{f'(R_0)-1}\equiv4{\Lambda_{\rm f}}<0.
\end{eqnarray}
Substituting the above relation into Eq. (\ref{FE1}), we obtain
the following equation for Ricci tensor
\begin{equation}
{R}_{\mu \nu}=\frac{1}{2}g_{\mu \nu}
\left(\frac{f(R_0)}{f^{\prime}(R_0)-1}
\right)+\frac{2}{1+f^{\prime}(R_0)}T_{\mu \nu}. \label{Ricci}
\end{equation}
Now, we want to construct charged rotating black string solutions
of the field equations (\ref{FE1}) and (\ref{FE2}) and investigate
their properties. We are looking for the four-dimensional rotating
solution with cylindrical or toroidal horizons. The metric which
describes such a spacetime can be written in the following form
\cite{Lem,shey}
\begin{eqnarray}\label{metric}
ds^{2} &=&-N(r)\left( \Xi dt-a d\phi \right) ^{2}+%
r^{2}\left(\frac{a}{l^2} dt-\Xi d\phi \right) ^{2}+\frac{dr^{2}}{N(r)}+\frac{r^{2}}{l^{2}}dz^{2},  \nonumber \\
\Xi ^{2} &=&1+\frac{a^{2}}{l^{2}},
\end{eqnarray}
where $a$ is the rotation parameter. The function $N(r)$ should be
determined and $l$ has the dimension of length which is related to
the constant $\Lambda_{\rm f} $ by the relation
$l^{2}=-3/\Lambda_{\rm f} $. The two dimensional space,
$t$=constant and $r$ =constant, can be (i) the flat torus model
$T^2$ with topology $S^1 \times S^1$, and $0\leq \phi<2\pi$,
$0\leq z<2\pi l$, (ii) the standard cylindrical model with
topology $R\times S^1$, and $0\leq \phi<2\pi$, $-\infty<
z<\infty$,  and (iii) the infinite plane $R^2$ with $-\infty<
\phi<\infty$ and $-\infty< z<\infty$. We will focus upon (i) and
(ii). The Maxwell equation (\ref{FE2}) can be integrated
immediately to give
\begin{eqnarray}
F_{tr} &=&\frac{q\Xi }{r^2}, \nonumber
\label{Ftr} \\
F_{\phi r} &=&-\frac{a}{\Xi }F_{tr},
\end{eqnarray}
where $q$ is the charge parameter of the black string.
Substituting the Maxwell fields (\ref{Ftr}) as well as the metric
(\ref{metric}) in the field equation (\ref{FE1}) with constant
curvature, the non-vanishing independent components of the field
equations for $a=0$ reduce to
\begin{eqnarray}
&&\left(1+{\it f^{\prime}(R_0)}\right)\left(2r^4 \frac {d^2 N
(r)}{d r^2} +4 r^3 \frac {d N(r)}{d
r}+R_0r^4\right)-4q^2=0,\label{Eqtt}\\
&&\left(1+{\it f^{\prime}(R_0)}\right)\left(4 r^3 \frac {d N(r)}{d
r}+4r^2 N(r)+R_0r^4\right)+4q^2=0. \label{Eqzz}
\end{eqnarray}
One can easily show that the above equations have the following
solution
\begin{equation}\label{f(r)}
N(r)=-\frac{2m}{r}+\frac{q^2}{(1+f'(R_0))r^2}-\frac{R_0}{12}r^2 , \label{Nr}
\end{equation}
where $m$ is an integration constant which is related to the mass
of the string. One can also check that these solutions satisfy
equations (\ref{FE1})-(\ref{FE2}) in the rotating case where
$a\neq0$. It is apparent that this spacetime is similar with
asymptotically AdS black string. Indeed, with the following
replacement
\begin{eqnarray}\label{rep}
&&\frac{q^2}{\left(1+f^{\prime}(R_0)\right)} \rightarrow Q^2\\
&&\frac{R_0}{4}\rightarrow \Lambda
\end{eqnarray}
the solution reduces to the asymptotically AdS charged black
string for $\Lambda=-3/l^2$ \cite{Lem}. Next we study the physical
properties of the solutions. The Kretschmann scalar for this
solution is given by
\begin{eqnarray}
R_{\mu\nu \lambda \kappa }R^{\mu \nu \lambda \kappa
}=\frac{8}{3r^8(1+f'(R_0))^2}\left[r^2(\frac{1}{16}{R_0}^2r^6+18m^2)(1+f'(R_0))^2
-36mrq^2(1+f'(R_0))+21q^4\right].
\end{eqnarray}
\begin{figure}[tbp]
\epsfxsize=7cm  \centerline{\epsffile{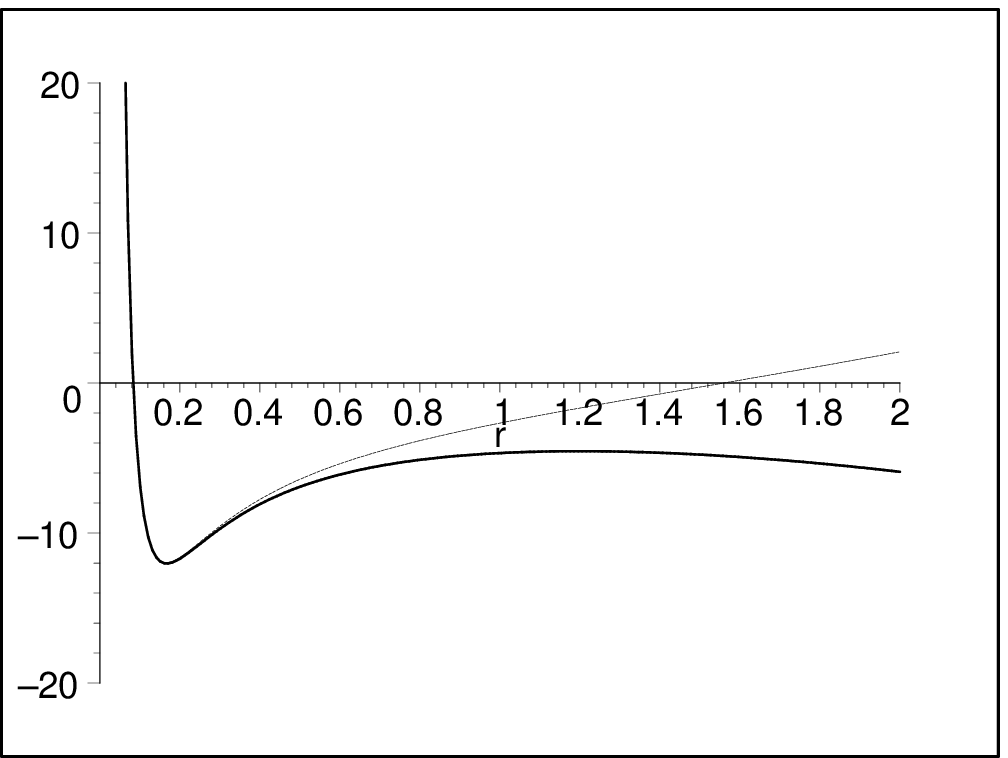}} \caption{The
function $N(r)$ versus $r$ for $m=2$, $f'(R_0)=2$ and $q=1$.
$R_0=12$ (bold line) and $R_0=-12$ (continuous line).}
\label{figure1}
\end{figure}
\begin{figure}[tbp]
\epsfxsize=7cm \centerline{\epsffile{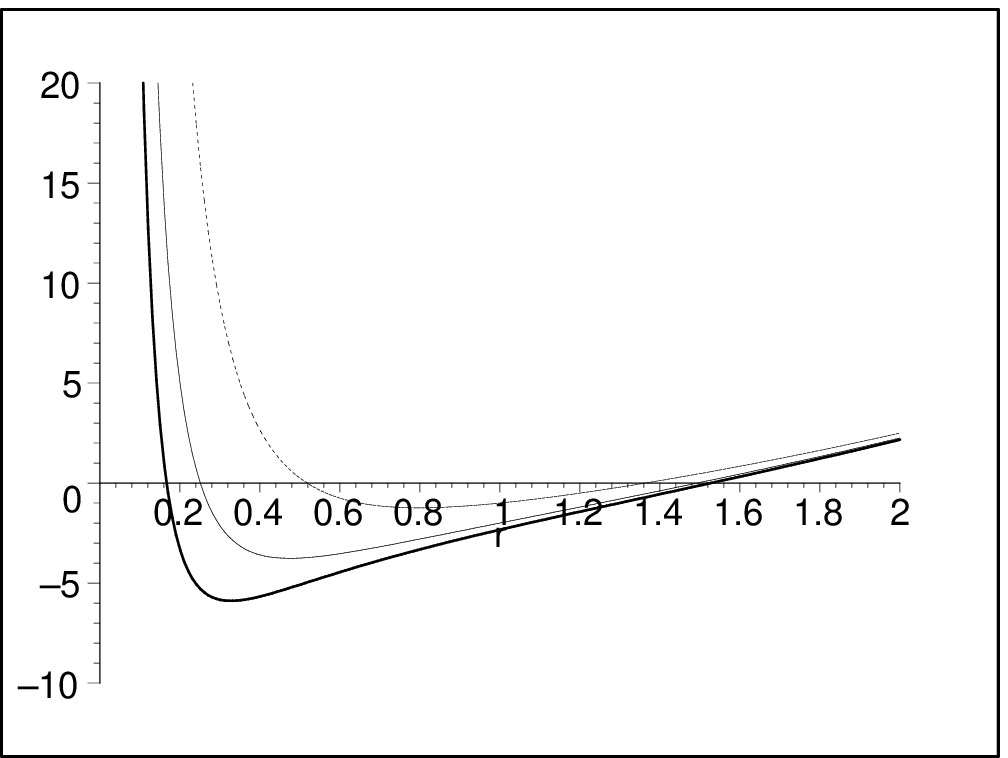}} \caption{The
function $N(r)$ versus $r$ for $m=2$, $q=1$ and $R_0=-12$.
$f'(R_0)=0.5$ (bold line), $f'(R_0)=0$ (continuous line) and
$f'(R_0)=-0.5$ (dashed line).} \label{figure2}
\end{figure}
\begin{figure}[tbp]
\epsfxsize=7cm  \centerline{\epsffile{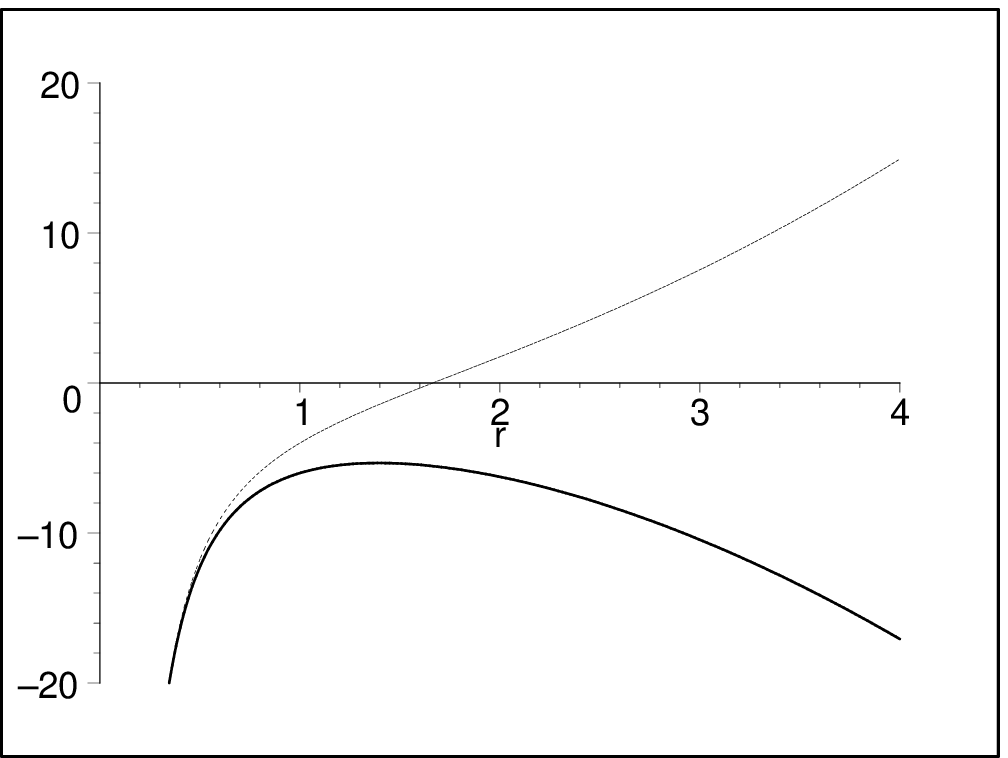}} \caption{The
function $N(r)$ versus $r$ for $m=2$, $f'(R_0)=-2$ and $q=1$.
$R_0=12$ (bold line) and $R_0=-12$ (continuous line).}
\label{figure3}
\end{figure}
When $r\rightarrow 0$, the dominant term in the Kretschmann scalar
is ${56q^4}/[(1+f'(R_0))^2r^8]$. Therefore we have an essential
singularity located at $r=0$. The Kretschmann scalar also
approaches $ {R_0}^2/6$ as $r\rightarrow \infty$. As one can see
from Eq. (\ref{f(r)}), the solution is ill-defined for
$f^{\prime}(R_0)=-1$. The cases with $f^{\prime}(R_0) >-1$ and
$f^{\prime}(R_0)<-1$ should be considered separately. In the first
case where $f^{\prime}(R_0) >-1$,  there exist a cosmological
horizon for $R_0>0$, while there is no cosmological horizons if
$R_0 <0$ (see fig. \ref{figure1}). Indeed, for
$1+f^{\prime}(R_0)>0$ and $R_0<0$ the black string can have two
inner and outer horizons provided the parameters of the solutions
are chosen suitably (see fig. \ref{figure2}). In the latter case
($f^{\prime}(R_0) <-1$), the signature of the spacetime changes
and the conserved quantities such as mass and angular momenta
become negative, as we will see in the next section, thus this is
not a physical case and we rule it out  from our consideration
(see fig. \ref{figure3} ).
\section{Conserved and Thermodynamic quantities\label{Therm}}
Next, we calculate the conserved quantities of the solutions by
using the counterterm method inspired by (A)dS/CFT correspondence
\cite{Mal}. The spacetimes under consideration in this paper has
zero curvature boundary, $R_{abcd}(h )=0$, and therefore the
counterterm for the stress energy tensor should be proportional to
$h^{ab}$. We find the suitable counterterm which removes the
divergences of the action in the form
\begin{equation}\label{cont}
 I_{\rm ct}=-\frac{1}{8\pi }\int_{\partial
 \mathcal{M}}d^{3}x\sqrt{-h
}\sqrt{-\frac{R_{0}}{3}},
\end{equation}
where $R_0<0$. Having the total finite action $I =
I_{G}+I_{\mathrm{ct}}$ at hand, one can use the quasilocal
definition  to construct a divergence free stress-energy tensor
\cite{BY}. Thus the finite stress-energy tensor can be written as
\begin{equation}
T^{ab}=\frac{1}{8\pi }\left[ \Theta ^{ab}-\Theta h
^{ab}-\sqrt{-\frac{R_{0}}{3}}h ^{ab}\right]. \label{Stres}
\end{equation}
The first two terms in Eq. (\ref{Stres}) are the variation of the
action (\ref{Act}) with respect to $h_{ab}$, and the last term is
the variation of the boundary counterterm (\ref{cont}) with
respect to $h_{ab}$. To compute the conserved charges of the
spacetime, one should choose a spacelike surface $ \mathcal{B}$ in
$\partial \mathcal{M}$ with metric $\sigma _{ij}$, and write the
boundary metric in ADM (Arnowitt-Deser-Misner) form:
\[
h_{ab}dx^{a}dx^{a}=-N^{2}dt^{2}+\sigma _{ij}\left( d\varphi
^{i}+V^{i}dt\right) \left( d\varphi ^{j}+V^{j}dt\right) ,
\]
where the coordinates $\varphi ^{i}$ are the angular variables
parameterizing the hypersurface of constant $r$ around the origin, and $N$
and $V^{i}$ are the lapse and shift functions respectively. When there is a
Killing vector field $\mathcal{\xi }$ on the boundary, then the quasilocal
conserved quantities associated with the stress tensors of Eq. (\ref{Stres})
can be written as
\begin{equation}
Q(\mathcal{\xi )}=\int_{\mathcal{B}}d^{2}x \sqrt{\sigma }T_{ab}n^{a}%
\mathcal{\xi }^{b},  \label{charge}
\end{equation}
where $\sigma $ is the determinant of the metric $\sigma _{ij}$, $\mathcal{%
\xi }$ and $n^{a}$ are, respectively, the Killing vector field and
the unit normal vector on the boundary $\mathcal{B}$. The first
Killing vector of the spacetime is $\xi =\partial /\partial t$,
and therefore its associated conserved charge of the string is the
mass per unit volume. A simple calculation gives
\begin{eqnarray}
M &=&\int_{\mathcal{B}}d^{2}x \sqrt{\sigma }T_{ab}n^{a}\xi
^{b}=\frac{(3\Xi ^{2}-1)m}{8\pi l}\left[1+f'(R_0)\right].
\label{M}
\end{eqnarray}
The second  conserved quantity is the angular momentum per unit
volume associated with the rotational Killing vectors $\varsigma
=\partial /\partial \phi$ which can be calculated as
\begin{eqnarray}
J =\int_{\mathcal{B}}d^{2}x \sqrt{\sigma }T_{ab}n^{a}\varsigma
^{b}= \frac{3\Xi
m\sqrt{\Xi^2-1}}{8\pi}\left[1+f^{\prime}(R_0)\right]. \label{J}
\end{eqnarray}
For $a=0$ ($\Xi =1$), the angular momentum per unit volume
vanishes, and therefore $a$ is the rotational parameters of the
spacetime. Next we calculate the entropy of the black string. Let
us first give a brief discussion regarding the entropy of the
black hole solutions in $f(R)$ gravity. To this aim, we follow the
arguments presented in \cite{Brevik}. If one use the Noether
charge method for evaluating the entropy associated with black
hole solutions in $f(R)$ theory with constant curvature, one finds
\cite{Cogn}
\begin{equation}
{S}=\frac{A}{4G}f^{\prime}(R_0), \label{S0}
\end{equation}
where $A=4 \pi r_{+}^2$ is the horizon area. As a result, in
$f(R)$ gravity, the entropy does not obey the area law and one
obtains a modification of the `` area law". Motivated by the above
argument, for the rotating black string solution in $R+f(R)$
gravity, we find the entropy per unit length of the string as
\begin{equation}
{S}=\frac{r_+^2\Xi}{4l} \left[1+f'(R_0)\right]. \label{S}
\end{equation}
Then we obtain the temperature and angular velocity of the horizon
by analytic continuation of the metric. Although our solution is
not static, the Killing vector
\begin{equation}\label{chi}
\chi=\partial_{t}+\Omega\partial_{\phi}
\end{equation}
is the null generator of the event horizon where $\Omega$
 is the angular velocity of the outer horizon. The
analytical continuation of the Lorentzian metric by $t\rightarrow
i\tau$ and $a\rightarrow ia$ yields the Euclidean section, whose
regularity at $r = r_+$ requires that we should identify $\tau\sim
\tau+\beta_+$ and $\phi\sim\phi+i\beta_+ \Omega_+$ where $\beta_+$
and $\Omega_+$ are the inverse Hawking temperature and the angular
velocity of the horizon. We find
\begin{eqnarray}\label{Tem}
T&=&\frac{1}{4\pi \Xi}\left(\frac{d N(r)}{dr}\right)_{r=r_{+}}
=-\frac{\left[R_0 r_{+}^4(1+f'(R_0))+4q^2
\right]}{16\pi\Xi[1+f'(R_0)] r_{+}^3},\\
\Omega&=&\frac{a}{\Xi l^2}, \label{Ome}
\end{eqnarray}
where we have used equation $N(r_{+})=0$ for omitting the mass
parameter $m$ from temperature expression. Since $1+f'(R_0)>0$,
therefore the temperature is non negative provided
\begin{eqnarray}\label{R0}
R_0 r_{+}^4(1+f'(R_0))\leq -4q^2 \rightarrow R_0 \leq
-\frac{4q^2}{r_{+}^4(1+f'(R_0))},
\end{eqnarray}
where the equality holds for extremal black string with zero
temperature. The next quantity we are going to calculate is the
electric charge of the string. To determine the electric field we
should consider the projections of the electromagnetic field
tensor on special hypersurface. The normal vectors to such
hypersurface are
\begin{equation}
u^{0}=\frac{1}{N},\text{ \ }u^{r}=0,\text{ \
}u^{i}=-\frac{V^{i}}{N},
\end{equation}
where $N$ and $V^{i}$ are the lapse function and shift vector.
Then the electric field is $E^{\mu }=g^{\mu \rho }F_{\rho \nu }u^{\nu }$,
and the electric charge per unit
length of the string can be found by calculating the flux of the
electric field at infinity,
\begin{equation}
{Q}=\frac{\Xi q}{4\pi l\sqrt{1+f'(R_0)}}.  \label{Q}
\end{equation}
The electric potential $U$, measured at infinity with respect to
the horizon, is defined by \cite{Cal}
\begin{equation}\label{U}
U=A_{\mu }\chi ^{\mu }\left| _{r\rightarrow \infty }-A_{\mu }\chi
^{\mu }\right| _{r=r_{+}},
\end{equation}
where $\chi$ is the null generator of the event horizon given in
Eq. (\ref{chi}). One can easily obtain the electric potential as
\begin{equation}
U=\frac{q}{\Xi r_{+}}\sqrt{1+f'(R_0)}. \label{Pott}
\end{equation}
Then, we consider the first law of thermodynamics for the black
string. In order to do this, we obtain the mass $M$ as a function
of extensive quantities $S$, ${J}$ and $Q$. Using the expression
for the mass, the angular momenta, the entropy and the charge
given in Eqs. (\ref{M}), (\ref{J}), (\ref{S}) and (\ref{Q}) and
the fact that $N(r_{+})=0$, one can obtain a Smarr-type formula as
\begin{equation}
M(S,J,Q)=\frac{J(3Z-1)}{3l\sqrt{Z(Z-1)}}, \label{Msmarr}
\end{equation}
where $Z=\Xi ^{2}$ is the positive real root of the following
equation:
\begin{eqnarray}
\frac{3\sqrt{Z-1}\pi^2Q^2l^2[1+f'(R_0)]^2-2J\pi\sqrt{\sqrt{Z}(1+f'(R_0))S
l} +3S^2\sqrt{Z-1}}{2\pi\sqrt{\sqrt{Z}(1+f'(R_0))S l}}=0.
\end{eqnarray}
One may then regard the parameters $S$, ${J}$ and $Q$ as a
complete set of extensive parameters for the mass $M(S,{J},Q)$ and
define the intensive parameters conjugate to $S$, ${J}$ and $Q$.
These quantities are the temperature, the angular velocities and
the electric potential
\begin{eqnarray} \label{inte1}
T&=&\left(\frac{\partial M}{\partial S}\right)_{J,Q}
=-\frac{J\left\{\left[\pi Q l(1+f'(R_0))\right]^2-3S^2\right\}}
{3Sl\sqrt{Z(Z-1)}\left\{\left[\pi Q l(1+f'(R_0))\right]^2+S^2\right\}}, \\
\Omega &=&\left(\frac{\partial M}{\partial J}\right)_{S,Q} \nonumber\\
&=&\frac{3\left(3Z-1\right)\left\{\left[\pi Q
l(1+f'(R_0))\right]^2+S^2\right\} -4\pi
J\sqrt{\sqrt{Z}(Z-1)(1+f'(R_0))S l}}
{9l\sqrt{Z(Z-1)}\left\{\left[\pi Q l(1+f'(R_0))\right]^2+S^2\right\}},\label{inte2} \\
U&=&\left(\frac{\partial M}{\partial Q}\right)_{S,J}
=\frac{4\pi^2QlJ[1+f'(R_0)]^2}{3\sqrt{Z(Z-1)}\left\{\left[\pi Q
l(1+f'(R_0))\right]^2+S^2\right\}}. \label{inte3}
\end{eqnarray}
Numerical calculations show that the intensive quantities
calculated by Eqs. (\ref{inte1})-(\ref{inte3}) coincide with Eqs.
(\ref{Tem}), (\ref{Ome}) and (\ref{Pott}), respectively. Thus,
these thermodynamics quantities satisfy the first law of
thermodynamics
\begin{equation}
dM=TdS+\Omega d{J}+Ud{Q}.
\end{equation}
\section{Thermal Stability of black string}\label{Stab}
\begin{figure}[bp]
\epsfxsize=7cm \centerline{\epsffile{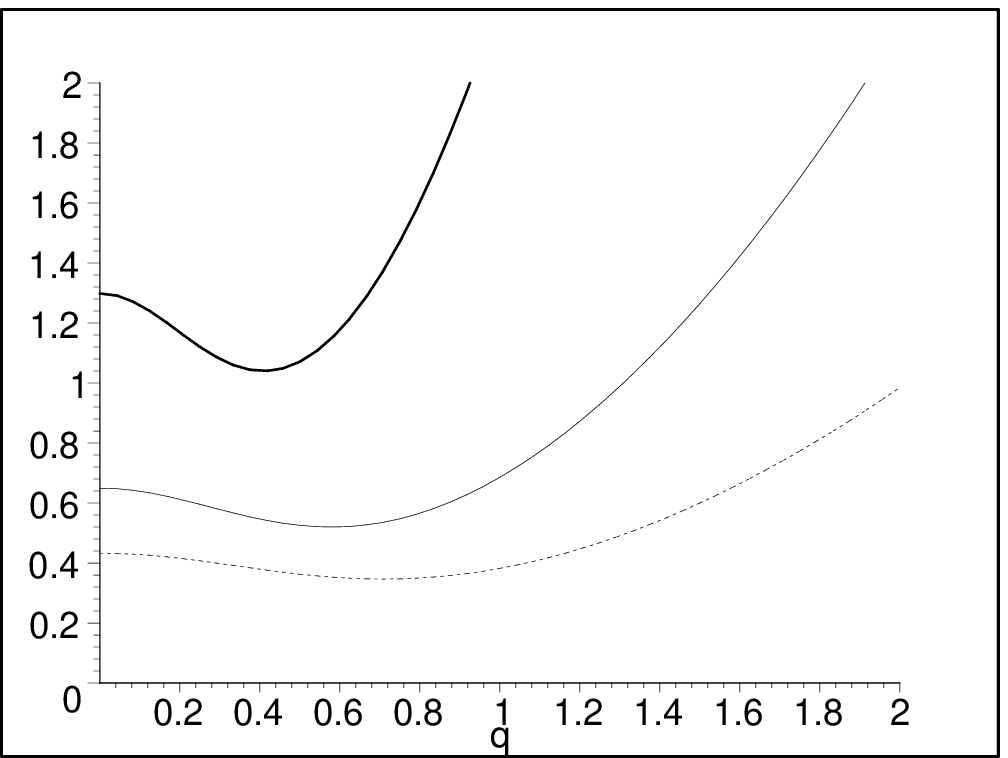}} \caption{The
function $(\partial ^2M/\partial S^2)_{J,Q}$ versus $q$ for $l=1$,
$\Xi=1.25$, $r_{+}=0.7$ and $R_0=-12$. $f'(R_0)=0$ (bold line),
$f'(R_0)=1$ (continuous line) and $f'(R_0)=2$ (dashed line).}
\label{figure4}
\end{figure}
\begin{figure}
\epsfxsize=7cm \centerline{\epsffile{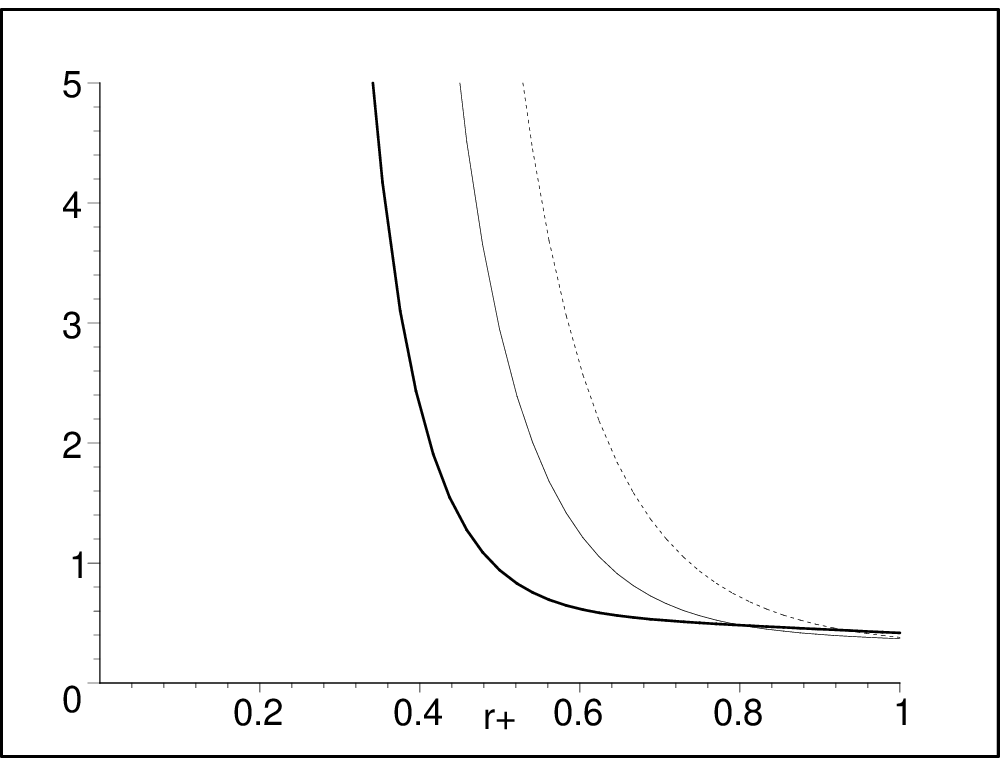}} \caption{The
function $(\partial ^2M/\partial S^2)_{J,Q}$ versus $r_{+}$ for
$l=1$, $\Xi=1.25$, $R_0=-12$ and $f'(R_0)=1$. $q=0.5$ (bold line),
$q=1$ (continuous line) and $q=1.5$ (dashed line).}
\label{figure5}
\end{figure}
\begin{figure}
\epsfxsize=7cm \centerline{\epsffile{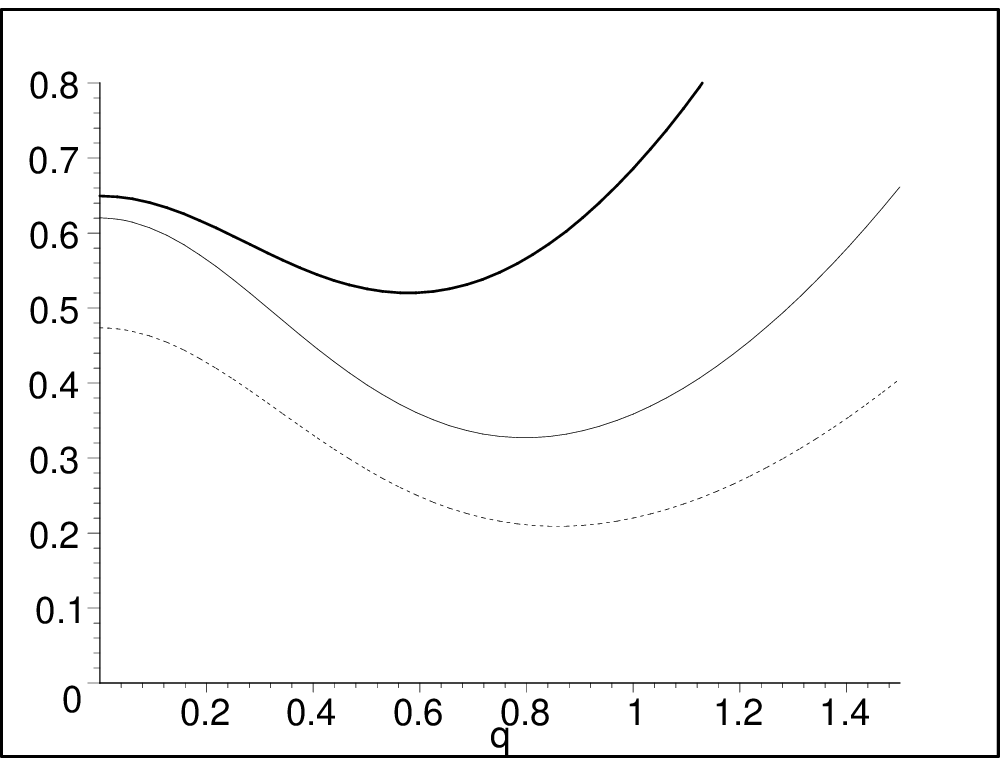}} \caption{The
function $(\partial ^2M/\partial S^2)_{J,Q}$ versus $q$ for $l=1$,
$f'(R_0)=1$ and $R_0=-12$. $\Xi=1.25$, (bold line), $\Xi=1.75$,
(continuous line) and $\Xi=2.25$, (dashed line).} \label{figure6}
\end{figure}
Finally, we investigate the thermal stability of rotating black
string solutions in  $f(R)$ gravity coupled to a matter field. The
stability of a thermodynamic system with respect to small
variations of the thermodynamic coordinates is usually performed
by analyzing the behavior of the entropy $S(M,J, Q)$ around the
equilibrium. The local stability in any ensemble requires that
$S(M,{J},Q)$ be a convex function of the extensive variables or
its Legendre transformation must be a concave function of the
intensive variables. The stability can also be studied by the
behavior of the energy $M(S,J, Q)$ which should be a convex
function of its extensive variables. Thus, the local stability can
in principle be carried out by finding the determinant of the
Hessian matrix of $M(S,{J},Q)$ with respect to its extensive
variables $X_{i}$, $\mathbf{H}_{X_{i}X_{j}}^{M}=[\partial
^{2}M/\partial X_{i}\partial X_{j}]$ \cite{Cal, Gub}. In our case
the mass $M$ is a function of entropy, angular momenta, and
charge. The number of thermodynamic variables depends on the
ensemble that is used. In the canonical ensemble, the charge and
the angular momenta are fixed parameters, and therefore the
positivity of the $(\partial ^{2}M/\partial S^{2})_{{J},Q}$ is
sufficient to ensure local stability. We find that the black
string solutions are always thermally stable independent of the
value of the parameters $q$ and $\Xi$. We have shown the behavior
of $(\partial ^{2}M/\partial S^{2})_{{J},Q}$\ as a function $q $
and $r_{+}$ for different value of $\Xi$ and $f'(R_0)$ in figures
4-6. These figures show that the black string solutions in
$f(R)$-Maxwell theory with constant  curvature scalar are always
thermally stable.
\section{Solution with non constant Ricci scalar}
In this section we would like to extend the study to the case
where the Ricci scalar is not a constant, instead we reconstruct
it as $R=R(r)$ as a result of our calculations. In this case, we
find out that we can obtain solution only in the absence of the
matter field. As we mentioned in the introduction, in general, the
field equations of $f(R)$ theory coupled to the matter field are
very complicated and hence it is not easy to find exact analytical
solutions. Thus we only consider the uncharged black string
solution. We start with the following action
\begin{equation}
S=\frac{1}{16 \pi}\int d^4 x\sqrt{-g}f(R)\,,\label{action}
\end{equation}
where $f(R)$ is a generic function of the Ricci scalar $R$. We
also modify our metric (\ref{metric}) a bit as follow
\begin{eqnarray}\label{metric2}
ds^{2} &=&-N(r)\mathrm{e}^{2\alpha(r)}\left( \Xi dt-a d\phi \right) ^{2}+%
r^{2}\left(\frac{a}{l^2} dt-\Xi d\phi \right)
^{2}+\frac{dr^{2}}{N(r)}+\frac{r^{2}}{l^{2}}dz^{2},
\end{eqnarray}
where we have added an additional function $\alpha(r)$ in the
metric coefficients. For simplicity we only consider the
non-rotating black string with $a=0$, thus the above metric
reduces to
\begin{eqnarray}\label{metric3}
ds^{2}=-N(r)\mathrm{e}^{2\alpha(r)}dt^{2}+\frac{dr^{2}}{N(r)}+
r^{2}d\phi ^{2}+\frac{r^{2}}{l^{2}}dz^{2}.
\end{eqnarray}
The scalar curvature for metric (\ref{metric3}) reads
\begin{eqnarray}
R  &=&
-3\,{\frac{dN\left(r\right)}{dr}}{\frac{d\alpha\left(r\right)}{dr}}
-2\,N\left(r\right)\left[{\frac{d}{dr}}
\alpha\left(r\right)\right]^{2}-{\frac{d^{2}N\left(r\right)}{d{r}^{2}}}
-2\,N\left(r\right){\frac{d^{2}\alpha\left(r\right)}{d{r}^{2}}}\nonumber\\
&&-\frac{4}{r}\frac{d N\left(r\right)}{dr}
-\frac{4N\left(r\right)}{r}\frac{d\alpha\left(r\right)}{dr}-\frac{2N\left(r\right)}{{r}^{2}}
\,.\label{R}
\end{eqnarray}
We use the Lagrangian multipliers method. In the framework of
Friedmann-Robertson-Walker universe this method was studied in
\cite{vile,Capozziello,Monica}, while for static spherically
symmetric black hole solutions it was investigated in \cite{Sebas,
Capozziello2}. In this approach one may consider the scalar
curvature $R$ as independent Lagrangian coordinates in addition to
the functions $\alpha(r)$ and $N(r)$, which appear from the metric
line element.

Introducing the Lagrangian multipliers $\lambda$, after using
(\ref{R}), the  action  (\ref{action}) can be written
\begin{eqnarray}
S &\equiv& \frac{1}{16 \pi}\int dt\int d{ r}\left(e^{\alpha(r)}r^2
\right)\left\{ f(R)-\lambda
\left\{R+\left\{3\,\left[{\frac{d}{dr}}N\left(r\right)\right]{\frac{d}{dr}}
\alpha\left(r\right)\right.\right. \right.    \nonumber\\
&& +2\,N\left(r\right)\left[{\frac{d}{dr}}
\alpha\left(r\right)\right]^{2}+{\frac{d^{2}N\left(r\right)}{d{r}^{2}}}
+2\,N\left(r\right)\frac{d^{2}\alpha\left(r\right)}{d{r}^{2}}+\frac{4}{r} \frac{dN\left(r\right)}{dr}\nonumber\\
&&+\frac{4N\left(r\right)}{r}\frac{d\alpha\left(r\right)}{dr}+{\frac{2N\left(r\right)}{{r}^{2}}}
\left.\left.\left.\right\}\right\}\right\}\,.
\end{eqnarray}
 Varying the above action with respect to $R$, one gets
\begin{equation}
 \lambda=f'(R),
\end{equation}
where the prime denotes the derivative with respect to the scalar
curvature $R$. Substituting this value and integrating by part,
the Lagrangian takes the form
\begin{eqnarray}
L(\alpha, d\alpha/dr, N, d N/dr, R, d
R/dr)&=&e^{\alpha}\left\{r^2\left[f(R)-R
f'(R)\right]-2f'(R)\left(r\frac{d N(r)}{dr}+N(r)\right)\right.
\nonumber\\
&& +\left.f''(R)\frac{d R}{d r}r^2\left(\frac{d N(r)}{d
r}+2N(r)\frac{d \alpha(r)}{dr}\right)\right\}\,.
\end{eqnarray}
Making the variation with respect to $\alpha$, one gets the first
equation of motion
\begin{eqnarray}
\label{one}&
&\frac{Rf'(R)-f(R)}{f'(R)}+\frac{2}{r^2}\left[N(r)+r\frac{d
N(r)}{dr}\right]\\\nonumber & &
+\frac{2N(r)f''(R)}{f'(R)}\left[\frac{d^2 R}{d
r^2}+\left(\frac{dN(r)/dr}{2 N(r)}\right)\frac{d R }{d
r}+\frac{f'''(R)}{f''(R)}\left(\frac{d R}{d
r}\right)^2\right]=0\,.
\end{eqnarray}
The variation with respect to $N(r)$ leads the second equation of
motion
\begin{equation}
\left[\frac{d\alpha(r)}{dr}\left(\frac{f''(R)}{f'(R)}\frac{d R}{d
r}\right)-\frac{f''(R)}{f'(R)}\frac{d^2 R}{d
r^2}-\frac{f'''(R)}{f'(R)}\left(\frac{d R}{d
r}\right)^2\right]=0\,,\label{two}
\end{equation}
while by making the variation with respect to $R$, we recover Eq.
(\ref{R}). Given $f(R)$, together with equation (\ref{R}), the
above equations form  a system of three differential equations for
the three unknown quantities $\alpha(r), N(r)$ and $R(r)$. We
would like to note that one advantage of this approach is that
$\alpha$ does not appear in Eq.(\ref{one}). In what follow, we
will find exact solutions of the above system of differential
equations.

In the special case of constant curvature $R=R_0$ and $\alpha=\rm
constant$, it is easy to show that the only solution of Eqs.
(\ref{R}) and(\ref{one}) is the Schwarzshild de Sitter black
string solution with flat horizon,
\begin{equation} \label{N2}
N(r)=-\frac{2m}{r}-\frac{\Lambda}{3}r^2,
\end{equation}
where $m$ is a constant of integration which can be interpreted as
the mass parameter of the black string and we have defined $
2\Lambda\equiv R_{0}-f(R_{0})/f'(R_{0})$ and $R_0=4\Lambda$.
Notice that in the absence of the matter field, $q=0$, solution
(\ref{Nr}) coincides with the result obtained in (\ref{N2}), as
expected.

Next, we consider the case of non constant Ricci curvature, but
still with $\alpha=\rm constant$.  From Eq. (\ref{two}) we have
\begin{equation}
f'''\left(\frac{d R}{dr}\right)^2+f''\left(\frac{d^2 R}{d
r^2}\right)=\frac{d^2}{dr^2}f'(R)=0,
\end{equation}
which has the following solution,
\begin{equation}
f'(R)=mr+n, \label{Fprime}
\end{equation}
where $m$ and $n$ are two integration constants. Given the
explicit form of $R$, we may  find $r$ as a function of Ricci
scalar and reconstruct $f'(R)$ realizing such solution. From Eq.
(\ref{R}) with constant $\alpha$, one gets
\begin{equation}
R =-{\frac{d^{2}N\left(r\right)}{d{r}^{2}}} -\frac{4}{r} \frac{d
N\left(r\right)}{dr} -\,{\frac{2N\left(r\right)}{{r}^{2}}}.
\label{R2}
\end{equation}
Using the fact that $(f''(R))d R/dr=d f'(R)/dr=m$ and $d f(R)/d
r=f'(R) d R/d r$, and multiplying Eq. (\ref{one}) by $f'(R)$, we
arrive at
\begin{equation}
-\frac{d^2 N(r)}{dr^2}\left(m+\frac{n}{r}\right)+\frac{4 m
N(r)}{r^2}+\frac{2n N(r)}{r^3}-\frac{m}{r}\frac{d N(r)}{d
r}=0\,.\label{equazione}
\end{equation}
When $m=0$, the solution of the above equation is ones obtained in
(\ref{N2}). For $n=0$, the general solution is
\begin{equation}
N(r)=-C_1 r^2+\frac{C_{2}}{r^2}. \label{L1}
\end{equation}
Substituting in Eq. (\ref{R2}) we again arrive at constant Ricci
scalar, $R=12 C_1$. Although in this case $f'(R)=df(R)/dR=mr$ is
not a constant, but still we have $df(R)/dr=0$, which implies that
$f(R)=\rm constant$. Next we look for the most general solution of
Eq.(\ref{equazione}) with  $n \neq 0$ and $m\neq 0$. Solving
(\ref{equazione}), we find
\begin{eqnarray}\label{N3}
N \left( r \right) =-C_1{r}^{2}+\frac {C_2}{r} \left[ 2
\,{n}^{3}-3mn^2r+6\,{m}^{2}{r}^{2}n-6{m}^{3} r^3\ln  \left(
m+\frac{n}{r} \right) \right],
\end{eqnarray}
where $C_1$ and $C_2$ are two arbitrary constants. Given solution
(\ref{N3}) one can basically construct $f(R)$ by using Eqs.
(\ref{Fprime}) and (\ref{R2}).  In order to simplify the above
solution, we choose $n=1$ and $C_2=-1/m$,
\begin{equation}
N \left( r \right) =3-\frac {2}{mr}-6\,{mr}-C_1{r}^{2}+6{m}^{2}
r^2\ln \left( m+\frac{1}{r} \right).
\end{equation}
For this general case the Ricci scalar becomes
\begin{equation} \label{R3}
{R(r)}=12C_1-72m^2\ln\left(m+\frac{1}{r}\right)+\frac{6\left(12m^3r^3+18m^2r^2+4mr-1\right)}{r^2
(mr+1)^2}.
\end{equation}
which is clearly not a constant. Now we want to reconstruct the
corresponding $f(R)$ theory. From Eq.(\ref{Fprime}), for $n=1$ one
has
\begin{equation}\label{fprim2}
f'(R)=\frac{d f(R)}{dR}=\frac{df(R)}{dr}\frac{dr}{dR}=mr+1.
\end{equation}
Integrating (\ref{fprim2}), by using (\ref{R3}), we get
\begin{equation}\label{fR2}
f[R(r)]=-
36m^2\ln\left(m+\frac{1}{r}\right)+\frac{3\left(12m^3r^3+18m^2r^2+4mr-2\right)}{r^2
(ar+1)^2}.
\end{equation}
Combining Eqs. (\ref{R3}), (\ref{fprim2}) and (\ref{fR2}), one
gets the following differential equation for function $f(R)$,
\begin{equation}\label{fR3}
\frac{3m^2}{[f'(R)]^2[f'(R)-1]^2}+f(R)-\frac{R}{2}+6C_1=0.
\end{equation}
This equation has a simple solution as
\begin{equation}\label{fR4}
f(R)=\frac{R}{2}-6C_1-48 m^2,
\end{equation}
but it has also another complicated solution which we have not
presented it here. Thus we have found a black string solution in
$f(R)$ gravity with non constant Ricci scalar. This approach also
leads to construct Ricci scalar as  a function of $r$, as given in
Eq. (\ref{R3}). The obtained solutions in this section differ from
that presented in \cite{Capozziello2} for axially symmetric
solutions in $f(R)$ gravity. It is worth mentioning that metric
(\ref{metric2}) has a good property for which its static and
rotating solutions coincide and so in this section we only study
the static case. Following the approach of this section, one can
easily check that solution (\ref{N3}) can be deduced for rotating
case where $a\neq0$. Besides, in this section we only considered
the case with $\alpha=\rm constant$, and derived the metric
function (\ref{N3}) as well as $R(r)$. The study can also be
generalized to the case where $\alpha=\alpha(r)$. We leave it and
also thermodynamic considerations of the obtained solution in this
section, for future investigations.

\section{Conclusions\label{Conc}}
In order to obtain the constant curvature black hole solution in
$f(R)$ gravity theory coupled to a matter field, the trace of the
energy-momentum tensor of the matter field should be zero
\cite{Moon}. Since the energy-momentum tensor of Maxwell field is
traceless in four dimensions, therefore spherically symmetric
black hole solutions from $f(R)$ theory coupled to Maxwell field
was derived in four dimensional spacetime \cite{Moon}.

In this paper we continued the study by constructing a new class
of charged rotating solutions in $f(R)$-Maxwell theory with
constant curvature scalar. This class of solutions describe the
four dimensional charged rotation black string with cylindrical or
toroidal horizons with zero curvature boundary. These solutions
are similar to asymptotically AdS black string of Einstein-Maxwell
gravity with suitably replacement of the parameters. However, the
solution presented in this paper has at least  two differences
from AdS black string solutions of Einstein-Maxwell gravity.
First, the conserved and thermodynamic quantities computed here
depend on function $f'(R_{0})$ and differ completely from those of
Einstein theory in AdS spaces. Clearly the presence of the general
function $f'(R_{0})$ changes the physical values of conserved and
thermodynamic quantities. Second, unlike Einstein gravity, the
entropy does not obey the area law for black string solutions in
$f(R)$-Maxwell theory as one can see from Eq (\ref{S}). We studied
the physical properties of the solutions and found a suitable
conterterm which removes the divergence of the action. We obtained
mass and angular momenta of the string through the use of
conterterm method. We also derived the entropy of the black string
in $f(R)$ gravity which has a modification from the area law. We
obtained a Smarr-type formula for the mass namely $M(S,J,Q)$ and
checked that the obtained conserved and thermodynamic quantities
satisfy the first law of black hole thermodynamics. Finally, we
explored the thermal stability of the solutions in the canonical
ensemble and showed that the black strings derived from $f(R)$-
Maxwell theory are always thermally stable. This is commensurate
with the fact that there is no Hawking-Page phase transition for
black objects with zero curvature horizon \cite{Wit}.

We also extend the study to the case where the Ricci scalar is not
a constant. For this purpose we used the Lagrangian multipliers
method and found an exact black string solution in $f(R)$ gravity.
In this approach one may consider the scalar curvature $R(r)$ as
an independent Lagrangian coordinates in addition to the metric
functions and deduce $R(r)$ as a solution of the field equations.
We found the explicit form of $R(r)$ as well as the metric
function which has a logaritmic term. It is worth noting that
since for the non constant Ricci scalar, the field equations of
$f(R)$ theory coupled to the matter field become very complicated,
in this  case, we could only derived analytical solution in the
absence of the matter field.

\acknowledgments{We thank the referee for constructive comments
which helped us to improve the paper significantly. We also
grateful to S. H. Hendi for useful comments and helpful
discussions. The work of A. Sheykhi has been supported financially
by Research Institute for Astronomy \& Astrophysics of Maragha
(RIAAM), Iran. A. Sheykhi also thank from the Research Council of
Shiraz University. }


\begin{thebibliography}{99}
\bibitem{Odin} S. Nojiri and S. D. Odintsov, Phys. Rev. D
{\bf74}, 086005 (2006);\\ S. Nojiri and S. D. Odintsov, Int. J.
Geom. Meth. Mod. Phys. {\bf4}, 115 (2007);\\  S. Nojiri and S. D.
Odintsov, J. Phys. Conf. Ser. {\bf66}, 012005 (2007);\\ S. Nojiri
and S. D. Odintsov, Phys. Rept.{\bf 505}, 59 (2011);\\ S.
Capozziello, S. Nojiri, S. D. Odintsov and A. Troisi, Phys. Lett. B
{\bf639}, 135 (2006);\\ S. Nojiri and S. D. Odintsov, Phys. Rev. D
{\bf78}, 046006 (2008).



\bibitem{Capo} S. Capozziello, V. F. Cardone, and A. Troisi, J. Cosmol.
Astropart. Phys. {\bf08}, 001 (2006);\\ S. Capozziello, V. F.
Cardone, and A. Troisi, Mon. Not. R. Astron. Soc. {\bf375}, 1423
(2007);\\ K. Atazadeh, M. Farhoudi and H. R. Sepangi, Phys. Lett. B
{\bf660}, 275 (2008);\\ C. Corda, Astropart. Phys. {\bf34}, 587
(2011).

\bibitem{Anto} A. De Felice, S. Tsujikawa, Living  Rev. Rel. {\bf13}  (2010) 3.
\bibitem{Nun} A. Nunez and S. Solganik, hep-th/0403159.
\bibitem{Fara} V. Faraoni, Phys. Rev. D {\bf74}, 104017 (2006).
\bibitem{Soti} T. P. Sotiriou and V. Faraoni, Rev. Mod. Phys. {\bf82}, 451 (2010).
\bibitem{Cruz} A. de la Cruz-Dombriz, A. Dobado and A. L. Maroto, Phys. Rev. D {\bf80}, 124011
(2009).
\bibitem{Psal} D. Psaltis, D. Perrodin, K. R. Dienes and I. Mocioiu, Phys. Rev. Lett. {\bf100}, 091101 (2008).
\bibitem{Nova} J. Novak, Phys. Rev. D {\bf58}, 064019 (1998).
\bibitem{Pun} C. S. J. Pun, Z. Kovacs and T. Harko, Phys. Rev. D {\bf78}, 024043 (2008).
\bibitem{Lobo} F. S. N. Lobo and M. A. Oliveira, Phys. Rev. D {\bf80}, 104012 (2009).
\bibitem{Cogn} G. Cognola, E. Elizalde, S. Nojiri, S. D. Odintsov and S. Zerbini, JCAP {\bf0502}, 010
(2005).
\bibitem{Sebas} L. Sebastiani and S. Zerbini, arXiv:1012.5230.
\bibitem{Moon} T. Moon, Y. S. Myung and E. J. Son,  Gen. Rel. Grav. {\bf43}, 3079
(2011), arXiv:1101.1153.
\bibitem{Habib} S. Habib Mazharimousavi,  M. Halilsoy, T. Tahamtan, Eur. Phys. J. C  {\bf72} (2012) 1958.
\bibitem{Alex} A. Larranaga, Pramana Journal of Physics {\bf78}, 697 (2012) ;\\ J. A. R. Cembranos, et al.,
arXiv:1109.4519.

\bibitem{Lem} J. P. S. Lemos, Class. Quantum Gravit. {\bf12}, 1081 (1995);\\
J. P. S. Lemos, Phys. Lett. B {\bf353}, 46 (1995).

\bibitem{shey} M. H. Dehghani, Phys. Rev. D {\bf66}, 044006 (2002);\\ M. H.
Dehghani and A. Khodam-Mohammadi, Phys. Rev. D {\bf67}, 084006
(2003);\\ A. M. Awad, Class. Quant. Grav. {\bf20}, 2827 (2003);\\
M. H. Dehghani and N. Farhangkhah, Phys. Rev. D {\bf71}, 044008
(2005);\\ A. Sheykhi, M. H. Dehghani, N. Riazi, and J. Pakravan,
Phys. Rev. D {\bf74}, 084016 (2006);\\ A. Sheykhi, Phys. Rev. D
{\bf78}, 064055 (2008).

\bibitem{Mal}  J. Maldacena, Adv. Theor. Math. Phys. {\bf 2}, 231
(1998);\\
E. Witten, Adv. Theor. Math. Phys. {\bf 2}, 253 (1998); \\ O.
Aharony, S. S. Gubser, J. Maldacena, H. Ooguri and Y. Oz, Phys.
Rep. {\bf 323}, 183 (2000).
\bibitem{BY}  J. D. Brown and J. W. York, Phys. Rev. D {\bf 47}, 1407 (1993).

\bibitem{Brevik} I. Brevik, S. Nojiri, S.D. Odintsov and L. Vanzo, Phys. Rev.
D {\bf70}, 043520  (2004).



\bibitem{Cal}  M. Cvetic and S. S. Gubser, J. High Energy Phys. {\bf 04},
024 (1999);\\ M. M. Caldarelli, G. Cognola and D. Klemm, Class.
Quant. Gravit. {\bf 17}, 399 (2000).
\bibitem{Gub}  S. S. Gubser and I. Mitra, J. High Energy Phys., \textbf{08},
018 (2001).

\bibitem{Wit} E. Witten, Adv. Theor. Math. Phys. \textbf{2}, 505 (1998).

\bibitem{vile} A. Vilenkin,
Phys. Rev. {\bf D 32} 2511 (1985).

\bibitem{Capozziello} S. Capozziello,  Int. J. Mod. Phys. D {\bf11}, 483, (2002)
\bibitem{Monica}
G.~Cognola, M.~Gastaldi and S.~Zerbini,
Int.\ J.\ Theor.\ Phys.\  {\bf 47}, 898 (2008).

\bibitem{Capozziello2} S. Capozziello and M. De Laurentis, Phys. Rept. {\bf509}, 167
(2011).

\end{thebibliography}
\end{document}